\documentclass[twoside,twocolumn]{article}

\usepackage{graphicx}

\usepackage[sc]{mathpazo} 
\usepackage[T1]{fontenc} 
\usepackage{microtype} 

\usepackage[english]{babel} 

\usepackage[hmarginratio=1:1,top=32mm,columnsep=20pt]{geometry} 
\usepackage[hang, small,labelfont=bf,up,textfont=it,up]{caption} 
\usepackage{booktabs} 

\usepackage{lettrine} 

\usepackage{enumitem} 
\setlist[itemize]{noitemsep} 

\usepackage{abstract} 

\usepackage{titlesec} 
\renewcommand\thesection{\Roman{section}} 
\renewcommand\thesubsection{\roman{subsection}} 
\titleformat{\section}[block]{\large\scshape\centering}{\thesection.}{1em}{} 
\titleformat{\subsection}[block]{\large}{\thesubsection.}{1em}{} 

\usepackage{fancyhdr} 
\usepackage{titling} 

\pretitle{\begin{center}\Huge\bfseries} 
\posttitle{\end{center}} 
\title{A rapid and accurate method of finding light leaks in photomultiplier
systems} 
\author{%
\textsc{John McMillan} \\[1ex] 
\normalsize Department of Physics and Astronomy, The University of
Sheffield\\ Sheffield, South Yorkshire, S3 7RH, Great Britain \\ 
\normalsize {j.e.mcmillan@sheffield.ac.uk} 
}
\date{\today} 
\renewcommand{\maketitlehookd}{%
\begin{abstract}
\noindent A rapid method of finding light leaks in photomultiplier
systems is described, in which an audible signal derived from the light
level is produced.  It uses equipment commonly available in
laboratories. In practice it is like using a geiger counter to detect
radioactivity.
\end{abstract}
}

\begin{document}

\maketitle

\noindent{\it Keywords}: photomultipliers, scintillation detectors,
cherenkov detectors

\section{Introduction}
When constructing experimental apparatus containing photomultipliers
or other photodetectors sensitive at the single photon level,
light leaks are often encountered as it is actually quite difficult to
fabricate a completely light tight enclosure.
These light leaks may be caused by inadequate connectors,
misaligned seals, pinholes in materials or even
opaque materials being not quite as opaque as had been hoped
\cite{Wr17a,Le94a}.
All light leaks must be found and rectified before the apparatus
will operate correctly.

\section{Standard technique}
The standard technique for detecting light leaks is to
connect the anode output of the photomultiplier to an oscilloscope
and observe the signal with the timebase set so that some hundreds
of microseconds are visible.  If the signal has a ``grassy'' look
with many pulses visible, then a light leak can be suspected.
If this signal changes as the room lights are switched on or off,
then a light leak is probable.

This is essentially the technique
described by Leo~\cite{Le94a}, who says that signals from
detectors with light leaks appear ``unusually intense and
noisy''.  This advice was valid in the days of analog oscilloscopes,
where a change of event rate resulted in a change in the brightness
of the trace.  With modern digital oscilloscopes, rate changes are
not so obvious.

A better approach is to use a photon counting technique.
The photomultiplier output is coupled to a
discriminator set to trigger at the single photon level.  The output
from this is then counted over a suitable period (one or ten seconds,
possibly more)
with a counter-timer module.
The counts rate should have a Poisson
distribution,  so successive counts should differ from each other
by typically the square root of the count.  If switching the room
lights affects this count rate, a light leak is certain.

Identifying the exact point in the apparatus at which the light leak occurs
is more of a problem.  Many hours can be spent with torches, black tape and
blackout cloth, alternately covering and uncovering areas of the
detector system in the search for pinhole leaks.  At
each point, the count period must elapse and the count be visually checked
against the previous, with the square root error computed by mental
arithmetic.  The experimenter must constantly switch from looking at the
apparatus and observing the readout of the ratemeter.

\section{Aural technique}
A more rapid aural technique is suggested which can easily be implemented
with standard units commonly available in laboratories.
It is modelled on the use of a sounder in a traditional geiger
counter.
As before, the output from the photomultiplier is coupled to a
discriminator set to trigger at the single photon level.
This is passed to a gate-delay generator and the individual
discriminator output pulses
are widened, to render them audible.
\begin{figure}
   \center{
   \includegraphics[scale=0.95]{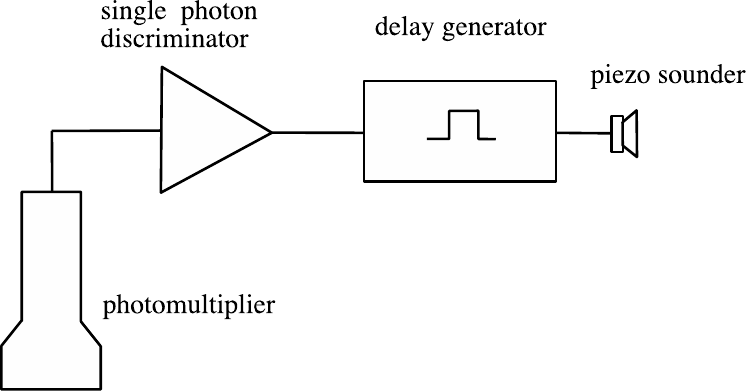}
          }
\caption{Aural light leak detection arrangement\label{block}}
\end{figure}
The gate-delay generator should be
chosen such that it has a TTL-output.  This output is fed to an audio
annunciator which can easily
be made by soldering a piezoelectric sounder or loudspeaker element
to a connector with a short coaxial cable.
The arrangement is shown in figure~\ref{block}.
A suitable sounder is the Pro-Signal ABI-003-RC which is widely
available, alternatively, piezo sounders scavenged from defunct
computers or smoke detectors also work.
If a suitable gate-delay generator is not available, a simple circuit
using a TTL monostable such as the 74LS221 can easily be constructed.

In use, the speaker produces a rushing sound,
the pitch of which is proportional to the
light level.  As the intensity of the light changes, so the sound
alters.  The effect is similar to using a geiger counter to detect radioactivity.
 In order to optimise performance, the photomultiplier should be
operated with minimum signal present.  This clearly depends on the
exact experimental setup.

By working in darkness and illuminating the apparatus with
blue or violet LED torches, or even laser pointers, the exact location
of leaks can be found in a small fraction of the time that the standard
technique takes.  The user does not have to wait for time periods to
elapse or to continuously observe
an oscilloscope or counter-timer display.

The choice of pulse width is determined experimentally.
The author has found that with small photomultipliers, $500\mu$s is a
good starting point.  If too short a value is chosen, then the sound
will be beyond human hearing, while if too long a value is chosen,
then many pulses are missed.

Note should be taken that with severe light leaks,
the illumination may be intense enough to saturate the
photomultiplier,
or the event rate might become so high that base-line shift occurs
and the single-photon discriminator no longer triggers.  In both cases
the sound stops.  Cutting the illumination should restart the sound.

After the light leak has been located and repaired, the photon
counting technique using the counter timer should be repeated with a
long count period to ensure that the detector system is completely
light tight.

\bibliographystyle{unsrt}
\bibliography{lightleaks}

\end{document}